\documentclass[12pt]{article}
\usepackage{epsfig}
\usepackage{url}
\usepackage{booktabs}

\def\beq{\begin{equation}}
\def\eeq#1{\label{#1}\end{equation}}
\def\eeqn{\end{equation}}

\def\beqa{\begin{eqnarray}}
\def\eeqa#1{\label{#1}\end{eqnarray}}
\def\eeqan{\end{eqnarray}}

\let\bar=\overbar

\def\Dslash{\not{\hbox{\kern-4pt $D$}}}
\def\dslash{\not{\hbox{\kern-2pt $\del$}}}

\def\BR{\mbox{\rm BR}}

\def\msb{{\bar{\ssstyle M \kern -1pt S}}}

\input pubboard/babarsym
\newcommand{\rhoz}{\ensuremath{\rho^0}}
\newcommand{\rhop}{\ensuremath{\rho^+}}
\newcommand{\rhom}{\ensuremath{\rho^-}}
\newcommand{\calB}{\ensuremath{{\cal B}}}
\newcommand{\timesix}{\ensuremath{\times10^{-6}}}
\newcommand{\etaprp}{\ensuremath{\eta^{(\prime)}}} 
\newcommand{\fetapiz}{\ensuremath{\eta\piz}\xspace}
\newcommand{\etapiz}{\ensuremath{\Bz\ra\fetapiz}\xspace}
\newcommand{\Betapiz}{\ensuremath{\calB(\etapiz)}\xspace}
\newcommand{\fetapeta}{\ensuremath{\etapr\eta}}
\newcommand{\etapeta}{\ensuremath{\Bz\ra\fetapeta}}
\newcommand{\Betapeta}{\ensuremath{\calB(\etapeta)}}

\newcommand{\Betappiz}{\ensuremath{\calB(\Bz\ra\etapr \piz)}\xspace}
\newcommand{\acp}{\ensuremath{\calA_{ch}}}

\def\Title#1{\begin{center} {\Large {\bf #1} } \end{center}}

\begin{document}

\begin{flushright}
  SLAC-PUB-12300 \\
  \babar-PROC-06-158 \\
  January 2007
\end{flushright}

\Title{Charmless \boldmath $B$ \unboldmath Decays}

\begin{center}{\large \bf Contribution to the proceedings of HQL06,\\
Munich, October 16th-20th 2006}\end{center}

\bigskip\bigskip


\begin{raggedright}  

{\it Wolfgang Gradl\index{Gradl, W.} (from the \babar{} Collaboration)\\
The University of Edinburgh\\
School of Physics\\
Edinburgh EH9 3JZ, UK}
\bigskip\bigskip
\end{raggedright}

\section{Introduction}
Rare charmless hadronic $B$ decays are a good testing ground for the
standard model.  
The dominant amplitudes contributing to this class of $B$ decays
are CKM suppressed tree diagrams and $b\to s$ or $b\to d$ loop
diagrams (`penguins').  These decays can be used to study interfering
standard model (SM) amplitudes and CP violation.  They are sensitive
to the presence of new particles in the loops, and they provide
valuable information to constrain theoretical models of $B$ decays.

The $B$ factories \babar{} at SLAC and Belle at KEK produce $B$
mesons in the reaction $\epem \to \FourS \to \BB$.  So far they have
collected integrated luminosities of about $406\invfb$ and
$600\invfb$, respectively.  The results presented here are based on
subsets of about $200$--$500 \invfb$ and are preliminary unless a
journal reference is given.

\section{\boldmath$\Delta S$ from rare decays\unboldmath}

The time-dependent CP asymmetry in $B$ decays is observed as an
asymmetry between \Bz{} and \Bzb{} decay rates into CP eigenstates $f$
\begin{equation}
  \label{eq:cpv}
          \mathcal{A}_{cp} (\Delta t) =  \frac{\Gamma(\Bzb\to f) -
            \Gamma(\Bz \to f)}{\Gamma(\Bzb\to f) + \Gamma(\Bz \to f)}
          = S_f \sin\Delta m_d \Delta t - C_f \cos\Delta m_d \Delta t
          , \nonumber
\end{equation}
where $\Delta m_d = 0.502\pm0.007\;\mathrm{ps^{-1}}$ and $\Delta t$ is
the time difference between the decays of the two neutral $B$ mesons
in the event.  The coefficients $S_f$ and $C_f$ depend on the final
state $f$; for the `golden' decay $\Bz\to\jpsi\KS$, for example, which
proceeds via a $b\to c\bar{c}s$ transition, only one weak phase
enters, and $S_{\jpsi\KS} = \sin2\beta$, $C_{\jpsi\KS} = 0$.  Here, $\beta \equiv
\phi_1$ is one of the angles of the unitarity triangle of the CKM
matrix.  

\begin{figure}[htb]
\begin{center}
\epsfig{file=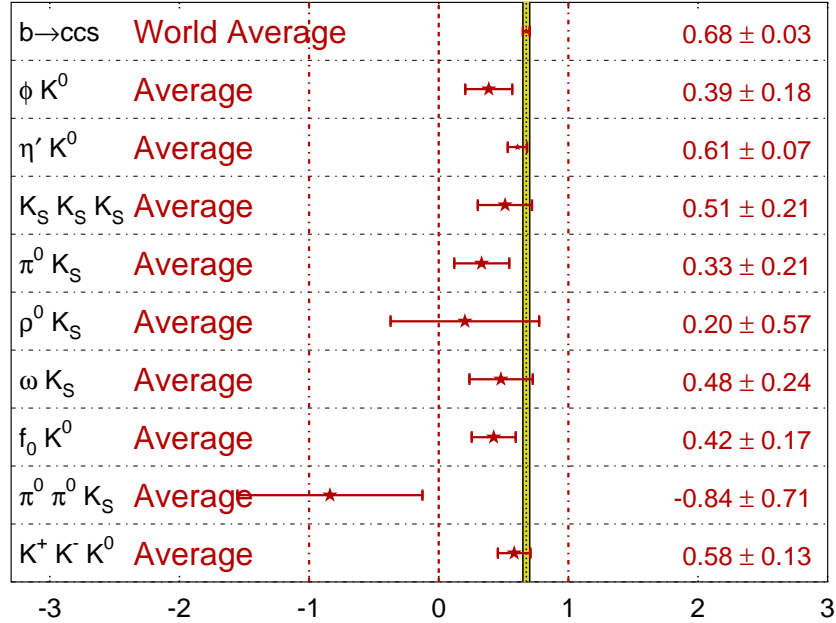,width=0.75\textwidth}
\caption{Average of $S_f$ in the different $b\to q\bar{q}s$ decays
  \cite{hfag1006}.}
\label{fig:S_in_qqs}
\end{center}
\end{figure}

In general, the presence of more than one contributing
amplitude for the decay can introduce additional phases, so that $S_f$
measured in such a decay deviates from the simple $\sin2\beta$.
Apart from standard-model amplitudes, particles beyond the standard
model may contribute in loop diagrams.
There
are intriguing hints in experimental data that $S_f$ is smaller than
$\sin 2\beta$ in $B$ decays involving the transition $b\to\qqbar s$,
like $\Bz\to\phi\Kz$, $\Bz\to\etapr\Kz$, or $\Bz\to\piz\Kz$ (see
Fig.~\ref{fig:S_in_qqs}).  However, 
for each of these final states the SM contribution to $\Delta
S_f\equiv S_f - \sin2\beta$ from sub-dominant amplitudes needs to be
determined in order to draw a conclusion about the presence of any new
physics.  Typically, models prefer $\Delta S_f > 0$
\cite{Beneke:2005pu,Cheng:2005bg}, while for the final state
$\etapr\KS$, a small, negative $\Delta S_f$ is
expected \cite{Williamson:2006hb}.  Measuring $B$ decays which are
related to the ones above by approximate SU(3) flavour or isospin
symmetries helps to constrain the standard-model expectation for 
$\Delta S_f$.

\subsection{\boldmath${\Bz\to\phi\Kz}$\unboldmath}
The sub-dominant amplitudes contributing to $\Bz\to\phi\Kz$ can be
constrained using SU(3) flavor relations \cite{Grossman:2003qp}.  This requires
branching fraction measurements for eleven decay channels
($\Kstarz\Kzb, \Kstarzb\Kz$, and $hh^\prime$ with $h = \rhoz, \omega,
\phi$ and $h^\prime = \piz, \eta, \etapr$).  \babar{} has measured an
upper limit for the sum $\calB(\Kstarz\Kzb) +
\calB(\Kstarzb\Kz) < 1.9\timesix$ \cite{Aubert:2006wu} and an updated upper
limit for $\phi\piz$ of $\calB(\phi\piz) < 2.8\times 10^{-7}$
\cite{Aubert:2006nn}.  Together with the already known upper limits or 
branching fractions for the other decays in this list, this allows one
to place a bound on $|\Delta S_{\phi\Kz}| < 0.43$ \cite{Aubert:2006wu}.

\subsection{\boldmath${\Betaprkz}$\unboldmath}
The decays $\Bz\to\etaprp\piz, \etapr\eta$ can be used to constrain
the SM pollution in \Betaprkz, The expected branching fractions are
between $0.2$ and $1 \times 10^{-6}$ for \etaprp\piz{} and $0.3$ -
$2\times 10^{-6}$ for $\etapr\eta$. Using 211\invfb{} of data,
\babar{} sets the following upper limits \cite{Aubert:2006qd} at 90\%
confidence level (C.L.)  in units of $10^{-6}$: $\Betapiz < 1.3$,
$\Betapeta < 1.7$, $\Betappiz <2.1$, while Belle \cite{Schumann:2006bg}
measures $\Betappiz = (2.79 ^{+ 1.02 + 0.25} _ {-0.96 -0.34})\timesix$
with $386\times 10^6$ analysed $\BB$ pairs.  With these new upper
limits, the standard model expectation for $\Delta S_{\etapr \KS}$ is
$-0.046 < \Delta S_{\etapr \KS} < 0.094$ \cite{Gronau:2006qh}.
A similar improvement for
the measurement of $\sin2\alpha$ in $\Bz\to\pip\pim$ is expected.  Belle also
measure $\calB(\Bp\to\etapr\pip) = (1.76 ^{+0.67 +0.15} _{-0.62
  -0.14})\timesix$ and a charge asymmetry in this channel of $\acp =
0.20 ^{+0.37} _{-0.36} \pm 0.04$.

\subsection{Pure penguin decays}
There is special interest in decays which only proceed via the $b\to
s\bar{s}s$ penguin transitions.  The $b\to u$ amplitudes can only
contribute through rescattering.   This drastically reduces the
standard model `pollution' in these decays, making them a very clean
probe for the presence of new particles in the loop.  An example for
this class of decays is $\Bz\to\KS\KS\KS$, in which the CP violating
parameters $S$ and $C$ have been measured by both \babar{}
\cite{hep-ex/0607108} and Belle \cite{hep-ex/0608039}, with an average
of $S = 0.51 \pm 0.21$, $C = -0.23 \pm 0.15$.
\babar{} has also searched for the related decay $\Bz\to\KS\KS\KL$.
The non-resonant contribution (besides $\Bz\to\phi(\KS\KL)\KS$) to
this final state has not been studied before and might be large
\cite{Cheng:2005ug}.  
Assuming a uniform Dalitz distribution and
analysing $211 \invfb$, \babar \cite{Aubert:2006zy} sets a 90\% CL
upper limit of $\calB(\Bz \to\KS\KS\KL) < 7.4\timesix$.  Due to a low
product of efficiency and daughter branching fraction, this decay is
therefore of limited use for the understanding of CP violation in
$b\to\qqbar s$ decays. 

\section{\boldmath Measurements related to $\alpha$}
Decays containing a $b\to u$ transition can be used to measure the
angle $\alpha \equiv \phi_2$ in the unitarity triangle. In general
several amplitudes with different weak phases contribute to these
decays, only allowing the direct measurement of an effective parameter
$\alpha_\mathrm{eff}$. There are several methods to extract the true
angle $\alpha$ in presence of this `pollution.'  Updated results for
the decays $B\to\rho\rho$, have been presented by Christos Touramanis
at this conference.

Another new decay studied by \babar{} and Belle is $\Bz\to
a_1^\pm\pi^\mp$, from which $\alpha$ can be extracted up to a
four-fold ambiguity.  Exploiting isospin or approximate SU(3) flavor
symmetries this ambiguity can be overcome \cite{Gronau:2005kw}.  
This needs also the
measurement of related axial--vector decays, from which a
model-dependent measurement of $\alpha$ can be derived.  \babar{}
searches for $\Bz\to a_1^\pm\pi^\mp$ in $211\invfb$ and
measures \cite{Aubert:2006dd} a branching fraction of $\calB(\Bz\to
a_1^\pm\pi^\mp) = (33.2\pm3.8\pm3.0)\timesix$, assuming $\calB(a_1^+
\to (3\pi)^+) = 1$.  With about the same luminosity, Belle measures a slightly larger branching
fraction of $(48.6 \pm 4.1 \pm 3.9)\timesix$  \cite{Abe:2005rf}. The
next step is to extend this analysis to measure time-dependent CP
violation in this decay.

\babar{} also searched for the related decay $\Bz \to a_1^+ \rhom$,
which also could be used to measure $\alpha$.  In addition, $B$ decays
to $5 \pi$ are an important background for the $B\to\rho\rho$
analyses.  In $100 \invfb$ no significant signal was seen; assuming a
fully longitudinal polarisation, the analysis sets a 90\% C.L. upper limit of
$\calB(\Bz\to a_1^+\rhom) \calB(a_1^+ \to (3\pi)^+) < 61\timesix$ \cite{Phys.Rev.D74.031104}.

\section{Charmless vector-vector decays}

For tree-dominated $B$ decays into two vector mesons, helicity
conservation arguments together with factorisation suggest that the
longitudinal polarisation fraction $f_L$ is $f_L \sim 1-m_V^2/m_B^2$,
close to unity.  Experimentally, this is seen in decays such as $B
\to \rho \rho$, where $f_L \approx 0.95$ is observed.  However, there
seems to be a pattern emerging where $f_L$ is smaller than the
expectation  in decays dominated by loop 
diagrams.  This was first seen in the decays
$B\to\phi\Kstar$, where $f_L$ is near $0.5$ with an uncertainty of
about 0.04 \cite{Phys.Rev.Lett.93.231804,Phys.Rev.Lett.94.221804}.

In the following sections, we describe a number of recent \babar{}
measurements for several of these vector-vector decays.  

\subsection{\boldmath Decays involving an $\omega$ meson}

To establish whether tree-induced decays generally have a large $f_L$,
\babar{} has searched for the related decays $B\to\omega V$ \cite{Aubert:2006vt},
where $V = \rho, \Kstar, \omega, \phi$.   The results are summarised
in Table~\ref{tab:omegaV}.  The only decay with a
significant observed yield is $\Bp\to\omega\rhop$ with
$\calB(\Bp\to\omega\rhop) = (10.6 \pm 2.1 ^{+1.6} 
_{-1.0})\timesix$.  The polarisation $f_L$ is floated in the fit and a
large value of $f_L = 0.82 \pm 0.11 \pm 0.02$ is found, as expected
for a tree-dominated decay. 

\begin{table}[b]
  \begin{center}
    \small 
    \begin{tabular}{lccccc} 
      \toprule
       & $\calB (10^{-6}) $ & $S (\sigma)$ & $\calB$ U.L \timesix &
      $f_L$  & $\acp$ \\
      \midrule
      $\omega \Kstarz$ & $2.4\pm 1.1 \pm 0.7$ & $2.4$ & $4.2$ & $0.71^{+0.27} _{-0.24}$ & -- \\
      $\omega \Kstarp$ & $0.6 ^{+1.4 +1.1} _{-1.2 -0.9}$ & $0.4$ & $3.4$ & 0.7 fixed & -- \\ 
      $\omega \rhoz$   & $-0.6 \pm 0.7 ^{+0.8} _{-0.3}$ & $0.6$ & $1.5$ & 0.9 fixed & -- \\
      $\omega f_0 (980)$&$0.9 \pm 0.4 ^{+0.2} _{-0.1} $ & $2.8$ & $1.5$ & -- & -- \\
      $\omega \rhop$   & $10.6 \pm 2.1 ^{+1.6} _{-1.0}$ & $5.7$ & -- & $0.82 \pm 0.11 \pm 0.02$ & $0.04 \pm 0.13 \pm 0.02$ \\
      $\omega \omega$  & $1.8 ^{+1.3} _{-0.9} \pm 0.4$  & $2.1$ & $4.0$ & $0.79 \pm 0.34$ & -- \\
      $\omega \phi$    & $0.1\pm 0.5 \pm 0.1$ & $0.3$ & $1.2$   & $0.88$ fixed & -- \\
      \bottomrule
    \end{tabular}
    \caption{Results of the \babar{} $\omega X$ analysis:  measured
      branching fraction $\calB$, significance including systematic
      uncertainties $S$, $90\%$ C.L. upper limit, measured or assumed
      longitudinal polarisation $f_L$, charge asymmetry $\acp$.}
    \label{tab:omegaV}
  \end{center}
\end{table}

\subsection{\boldmath $B \to \rho\Kstar$}

Conversely, the decays $B\to\rho\Kstar$ are penguin-dominated; some
are known to have significant branching fractions and $f_L$ can be
measured.  \babar{} has published updated measurements of branching
fractions, charge asymmetries and polarisation fractions
\cite{Aubert:2006fs}.

\subsubsection{\boldmath $\Bp \to \rhop\Kstarz$}
The decay $\Bp \to \rhop\Kstarz$ is particularly interesting because
no tree diagram is thought to contribute to this decay.  
\babar{} has a new measurement of the branching fraction, CP asymmetry
and polarisation for this decay.  The measured branching fraction is
$\calB(\rhop\Kstarz) = (9.6 \pm 1.7 \m 1.5)\timesix$,
$\acp(\rhop\Kstarz) = -0.01 \pm 0.16 \pm 0.02$.  The observed
polarisation is $f_L = 0.52 \pm 0.10 \pm 0.04$, as expected for a pure
penguin decay and in good agreement with $\phi \Kstar$.

\subsubsection{\boldmath $\Bp \to \rhoz\Kstarp$ and $\Bz \to \rhoz\Kstarz$ }
The decays $\Bp \to \rhoz\Kstarp$ and $\Bz \to \rhoz\Kstarz$ are
theoretically less clean because there is a Cabibbo-suppressed tree
diagram contributing in addition to the penguin present for all
$B\to\rho\Kstar$ decays.  In addition, $\Bp \to \rhoz\Kstarp$ is
experimentally more challenging because of the smaller branching
fraction.

For $\Bp \to \rhoz\Kstarp$, \babar{} measures a branching fraction of
$(3.6^{+1.7} _{-1.6} \pm 0.8) \timesix$, with a significance of only
$2.6\sigma$.  The value of $f_L$ determined by the fit is $f_L = 0.9
\pm 0.2$ although this is not considered a measurement for this decay,
as the signal itself is not significant.

$\Bz \to \rhoz\Kstarz$ is observed with a significance of $5.3\sigma$;
the branching fraction is $(5.6 \pm 0.9 \pm 1.3)\timesix$ and $f_L =
0.57 \pm 0.09 \pm 0.08$.

\section{\boldmath $B \to \etaprp K^{(*)}$}

In $B$ decays to final states comprising $\etaprp K^{(*)}$, the effect
of the $\eta$--$\etapr$ mixing angle combines with differing
interference in the penguin diagrams to suppress the final states
$\eta K$ and $\etapr \Kstar$, and enhance the final states $\etapr K$
and $\eta \Kstar$.   This pattern has now been experimentally
established with rather precise measurements of the branching
fractions for $\etapr K$  and $\eta \Kstar$ and the observation of the
decays $\etapr\Kstar$.  These decays are also important in light of
measuring $S$ in $\Bz \to \etapr \Kz$.

\subsection{\boldmath $B \to \etapr K$}
Belle's measurements for the branching fractions of
$B\to\etapr \pi$ \cite{Schumann:2006bg} were already mentioned above.
The same analysis also obtains updated branching fraction measurements
for the decays $B\to\etapr K$, with the results
$\calB(\Bz\to\etapr\Kz) = (58.9 ^{+3.6}_{-3.5} \pm 4.3) \timesix$, 
$\calB(\Bp\to\etapr\Kp) = (69.2 \pm 2.2 \pm 3.7) \timesix$, 
$\acp(\Bp\to\etapr\Kp) = 0.028 \pm 0.028 \pm 0.021$.

\subsection{\boldmath $B \to \eta \Kstar$ and $B \to \eta \rho$}
\babar{} \cite{hep-ex/0608034} and Belle \cite{hep-ex/0608005} have
published updated results for the decays $B \to \eta \Kstar(892)$.   Belle
also observes the decay $\Bp \to \eta\rhop$ and obtains an upper limit
for $\Bz \to eta\rhoz$.  These results confirm earlier measurements of
$B \to \eta \Kstar$ and $\eta\rho$.  \babar{} also analyses the mass region $1035
< m_{K\pi} < 1535\;\mathrm{MeV}$ of the $K\pi$ system and obtains
branching fractions for the spin-0  ($\eta (K\pi)^*_0$) and spin-2
($\eta \Kstar_2$) contributions. For these two final states no
predictions exist so far. The branching fraction results are
summarised in Table~\ref{tab:etaK}.

\begin{table}
  \centering
  \begin{tabular}{lcc}
    \toprule
    & \multicolumn{2}{c}{$\calB (10^{-6})$} \\
    & \babar & BELLE \\ 
    \midrule
    $\Bz \to \eta \Kstarz$ & $ 16.5 \pm 1.1 \pm 0.8 $ & $ 15.9 \pm 1.2 \pm 0.9 $         \\
    $\Bp \to \eta \Kstarp$ & $ 18.9 \pm 1.8 \pm 1.3 $ & $ 19.7 ^{+2.0} _{-1.9} \pm 1.4 $ \\
    $\Bp \to \eta \rhop$   & --                       & $ 4.1 ^{+1.4} _{-1.3} \pm 0.34 $ \\
    $\Bz \to \eta \rhoz$   & --                       & $ < 1.9 $                        \\
    \midrule
    $\Bz \to \eta (K\pi)_0^{*0} $ & $ 11.0 \pm 1.6 \pm 1.5 $  & --                       \\
    $\Bp \to \eta (K\pi)_0^{*+} $ & $ 18.2 \pm 2.6 \pm 2.6 $  & --                       \\
    $\Bz \to \eta  \Kstarz_2 $    & $ 9.6 \pm 1.8 \pm 1.1 $   & --                       \\
    $\Bp \to \eta  \Kstarp_2 $    & $ 9.1 \pm 2.7 \pm 1.4 $   & --                       \\
    \bottomrule
  \end{tabular}
  \caption{Branching fractions for the decays $B \to \eta \Kstar$, $\eta \rho$, and $\eta (K\pi)$. }
  \label{tab:etaK}
\end{table}

\subsection{\boldmath $B \to \etapr\Kstar$ and $B \to \etapr \rho$}
\babar{} \cite{Aubert:2006as} finds evidence for
the decays $B\to\etapr \Kstar$ in 211\invfb and measures branching
fractions of $\calB(\Bp\to\etapr\Kstarp) =
(4.9^{+1.9}_{-1.7}\pm0.8)\timesix$ and $\calB(\Bz\to\etapr\Kstarz) =
(3.8\pm1.1\pm0.5)\timesix$.  For the related decays into $\etapr\rho$,
only $\Bp\to\etapr\rhop$ is seen with $\calB(\Bp\to\etapr\rhop) =
(8.7^{+3.1}_{-2.8}{}^{+2.3}_{-1.3})\timesix$, while $\Bz\to\etapr
\rhoz$ is small with a 90\% C.L. upper limit of
$\calB(\Bz\to\etapr\rhoz) < 3.7\timesix$.  The direct CP asymmetries
in the decays with a significant signal are compatible with zero.
Theoretical predictions using SU(3) flavor
symmetry \cite{Chiang:2003pm}, QCD factorisation \cite{Beneke:2003zv},
and perturbative QCD factorisation \cite{Liu:2005mm} agree within
errors with the observed branching fractions.  The observation of
small branching fractions for $B \to \etapr \Kstar$ confirms the
pattern of enhanced and suppressed decays to $\etaprp K^{(*)}$.

\section{\boldmath $B \to \pi\pi, \pi K, KK$}
Updated branching fraction measurements for the two-body decays $B \to
\pi\pi, \pi K$, and $KK$ from \babar{}
\cite{hep-ex/0608003,hep-ex/0607106,hep-ex/0608036,hep-ex/0607096} and
Belle
\cite{hep-ex/0609015,Phys.Rev.Lett.94.181803,hep-ex/0608049,hep-ex/0609015}
are summarised in Table~\ref{tab:hh}.  Both experiments observe the
decays $\Bp \to \Kzb\Kp$ and $\Bz\to\Kzb\Kz$ with a statistical
significance $> 5\sigma$; decays with $b\to d$ hadronic penguins have
now been observed.

\babar{} also studied time dependent CP violation in $\Bz\to\Kzb\Kz$
\cite{hep-ex/0608036} (reconstructed as $\Bz\to\KS\KS$) which is a
pure $b\to d\bar{s}s$ penguin decay.  Via flavour SU(3) symmetry, this
decay also allows an estimate of the penguin contribution in
$\Bz\to\piz\piz$.  Direct CP asymmetry is expected to be zero.  The
result of the time-dependent fit is $S = -1.28
^{+0.80+0.11}_{-0.73-0.16}$ and $C = - 0.40 \pm 0.41 \pm 0.06$.

\begin{table}
  \centering
  \begin{tabular}{lccc}
    \toprule
     & \babar & BELLE \\
    & $\calB (10^{-6}) $ & $\calB (10^{-6})$ \\
    \midrule
    \pip\pim & $5.4\pm0.4\pm0.3$       & $5.1 \pm 0.2 \pm 0.2$              \\
    \Kp\pim  & $ 18.6 \pm 0.6 \pm 0.6 $& $20.0 \pm 0.4 ^{+0.9}_{-0.8}$      \\
    \Kp\Km   & $ < 0.40 $              &  ---                               \\
    \midrule
    $\Bz \to \piz\piz$ &  $1.48 \pm 0.26 \pm 0.12$ &  $2.3 ^{+0.4 +0.2} _{-0.5 -0.3}$ \\
    $\Bp \to \pip\piz$ &  $5.12 \pm 0.47 \pm 0.29$ &  $6.6 \pm 0.4 ^{+0.4}_{-0.5}$ \\
    $\Bpm\to \Kpm\piz$ &  $13.3 \pm 0.56 \pm 0.64$ &  $12.4 \pm 0.5 ^{+0.7}_{-0.6}$\\
    \midrule
    $\Bp \to \Kz \pip$ &  $23.9 \pm 1.1 \pm 1.0$   &  $22.9 ^{+0.8}_{-0.7}\pm1.3$ \\
    $\Bp \to \Kzb\Kp$&  $1.61 \pm 0.44 \pm 0.09$ &  $1.22 ^{+0.33 +0.13} _{-0.28 -0.16}$ \\
    $\Bz \to \Kzb\Kz$&  $1.08 \pm 0.28 \pm 0.11$ &  $0.86 ^{+0.24}_{-0.21} \pm 0.09$ \\
    \midrule
    $\Bz \to \KS \piz$ &  $10.5 \pm 0.7 \pm 0.5$   &  $9.2 ^{+0.7+0.6}_{-0.6-0.7}$ \\
    \bottomrule
  \end{tabular}
  \caption{Branching fraction results for $B \to \pi\pi,\pi K, K K$}
  \label{tab:hh}
\end{table}

\subsection{\boldmath $B \to \etapr\etapr K$, $\phi\phi K$}

Motivated by the large branching fraction for $B \to \etapr K$ and the
observation that final states $P^0P^0X^0$ are CP eigenstates
\cite{Gershon:2004tk}, \babar{} searched for the decays $B \to
\etapr\etapr K$.  No significant signal was found in $211\invfb$, and
the upper limits on the branching fractions of $\calB(\etapr\etapr\Kp)
< 25\timesix$ and $\calB(\etapr\etapr\Kz) < 31 \timesix$ are set
\cite{Aubert:2006ca}.

BELLE searched for the decays $B \to\phi\phi K$.  In these, direct CP
violation could be enhanced in the interference between decays via the
$\eta_c$ and non-SM decays.  In the analysis \cite{hep-ex/0609016},
charmless decays are selected by requiring that $m_{\phi\phi}$ is
below the charm threshold.  For these charmless decays, the observed
branching fractions are $\calB(\phi\phi\Kp) = (3.2 ^{+0.6} _{-0.5}\pm
0.3) \timesix$, $\calB(\phi\phi\Kz) = (2.3 ^{+1.0} _{-0.7}\pm
0.2) \timesix$.  The measured direct CP asymmetries are compatible
with zero.

\section{Summary}
Charmless hadronic $B$ decays provide a rich field for tests of QCD
and the standard model of electroweak interactions.  They allow to
constrain the SM contribution to $\Delta S_f$ in loop-dominated $B$
decays and precision tests of QCD models.  The $B$ factories have
produced a large number of new and updated measurements. With the
currently analysed statistics, decays with branching fractions of the
order of $10^{-6}$ are within experimental reach.

\bibliographystyle{h-physrev3}
\bibliography{references}

\end{document}